\documentclass[12pt]{article}

\usepackage[letterpaper, margin=1in]{geometry}

\usepackage[colorlinks]{hyperref}
\usepackage{makeidx}
\usepackage{framed}
\usepackage{fancyvrb}
\usepackage{color}
\usepackage{hyperref}
\hypersetup{colorlinks = true, allcolors = blue}
\usepackage{float}
\usepackage{flafter}
\usepackage{tabularx}
\usepackage{booktabs}
\usepackage{longtable}
\usepackage{multirow}
\usepackage{graphicx}
\usepackage{afterpage}

\usepackage{datetime}

\yyyymmdddate

  \usepackage{setspace}
  \AtBeginEnvironment{tabular}{\singlespacing}
  \AtBeginEnvironment{table}{\singlespacing}
  \AtBeginEnvironment{longtable}{\singlespacing}
  \AtBeginEnvironment{verbatim}{\singlespacing}
  \AtBeginEnvironment{Shaded}{\singlespacing}

\newlength{\cslhangindent}
\newenvironment{cslreferences}[2]
  {\setlength{\parindent}{0pt}
  \everypar{\setlength{\hangindent}{\cslhangindent}}\ignorespaces}
  {\par}

  \usepackage{amsmath}
  \usepackage{amssymb}

\usepackage{xcolor}
\definecolor{dark}{HTML}{2c2e35}
\color{dark}

\definecolor{myblue}{HTML}{1e3765}
\hypersetup{colorlinks,linkcolor=myblue,urlcolor=myblue}

\makeindex

\providecommand{\tightlist}{\setlength{\itemsep}{0pt}\setlength{\parskip}{0pt}}

  \usepackage{color}
  \usepackage{fancyvrb}

  \DefineVerbatimEnvironment{Highlighting}{Verbatim}{commandchars=\\\{\}}
  \usepackage{framed}
  \definecolor{shadecolor}{RGB}{248,248,248}
  \newenvironment{Shaded}{\begin{snugshade}}{\end{snugshade}}

  \newcommand{\CharTok}[1]{\textcolor[rgb]{0.31,0.60,0.02}{#1}}
  \newcommand{\CommentTok}[1]{\textcolor[rgb]{0.56,0.35,0.01}{\textit{#1}}}

  \newcommand{\ControlFlowTok}[1]{\textcolor[rgb]{0.13,0.29,0.53}{\textbf{#1}}}
  \newcommand{\DataTypeTok}[1]{\textcolor[rgb]{0.13,0.29,0.53}{#1}}
  \newcommand{\DecValTok}[1]{\textcolor[rgb]{0.00,0.00,0.81}{#1}}

  \newcommand{\FloatTok}[1]{\textcolor[rgb]{0.00,0.00,0.81}{#1}}

  \newcommand{\KeywordTok}[1]{\textcolor[rgb]{0.13,0.29,0.53}{\textbf{#1}}}
  \newcommand{\NormalTok}[1]{#1}
  \newcommand{\OperatorTok}[1]{\textcolor[rgb]{0.81,0.36,0.00}{\textbf{#1}}}
  \newcommand{\OtherTok}[1]{\textcolor[rgb]{0.56,0.35,0.01}{#1}}

  \newcommand{\StringTok}[1]{\textcolor[rgb]{0.31,0.60,0.02}{#1}}

\author{
  Mauricio Vargas Sepúlveda (ORCID 0000-0003-1017-7574)\\Department of
Political Science, University of Toronto\\Munk School of Global Affairs
and Public Policy, University of Toronto\\
  \smallskip\\
  Thomas J. Leeper (ORCID 0000-0003-4097-6326)\\
  \smallskip\\
  Tom Paskhalis (ORCID 0000-0001-9298-8850)\\Department of Political
Science, Trinity College Dublin\\
  \smallskip\\
  Manuel Aristarán\\
  \smallskip\\
  Jeremy B. Merrill\\The Washington Post\\
  \smallskip\\
  Mike Tigas (ORCID 0009-0008-7662-1619)\\
  Corresponding author: \href{mailto:m.sepulveda@mail.utoronto.ca}{\nolinkurl{m.sepulveda@mail.utoronto.ca}}
}
\title{tabulapdf: An R Package to Extract Tables from PDF Documents}
\date{Last updated: \today\ \currenttime}

\begin{document}

\maketitle

\thispagestyle{empty}
\tableofcontents
\setcounter{page}{0}
\clearpage

\afterpage{\setlength\parskip{10pt}}

\hypertarget{abstract}{%
\section{Abstract}\label{abstract}}

\texttt{tabulapdf} is an R package that utilizes the Tabula Java library
to import tables from PDF files directly into R. This tool can reduce
time and effort in data extraction processes in fields like
investigative journalism. It allows for automatic and manual table
extraction, the latter facilitated through a Shiny interface, enabling
manual areas selection with a computer mouse for data retrieval.

\hypertarget{introduction}{%
\section{Introduction}\label{introduction}}

For the user who needs raw data, PDF is not the right choice (James C.
King 2011, 2010). Even if a PDF file is created with a spreadsheet tool
such as Microsoft Excel, the resulting file does not contain information
about the data structure (i.e., rows and columns). The PDF format was
created as digital paper and one of its purposes was to share
information, this is, communicating interpretations and conclusions,
something different from sharing figures that usually belong in a
database (James C. King 2010; Merrill 2013).

Tabula is a multi-platform tool written in Java for extracting tables in
PDF files. For the previous reasons, extracting data provided in PDFs
can be challenging and time-consuming, and this tool allows to extract
tables into a CSV or Microsoft Excel spreadsheet using a simple,
easy-to-use interface.

One notable use case for Tabula is in investigative journalism, and it
was used to produced parts of the following stories (Aristaran 2018):

\begin{itemize}
\tightlist
\item
  \href{https://www.nytimes.com/interactive/2015/12/10/us/gun-sales-terrorism-obama-restrictions.html?_r=1}{The
  New York Times: ``What Happens After Calls for New Gun Restrictions?
  Sales Go Up''}
\item
  \href{https://foreignpolicy.com/2014/06/13/money-down-the-drain-crs-report-details-u-s-expenses-on-iraq/}{Foreign
  Policy: ``Money Down the Drain: CRS Report Details U.S. Expenses on
  Iraq''}
\item
  \href{https://projects.propublica.org/docdollars/}{ProPublica: How
  Industry Dollars Reached Your Doctors}
\end{itemize}

\texttt{tabulapdf} provides R bindings to the
\href{https://github.com/tabulapdf/tabula-java/}{Tabula java library},
which can be used to computationally extract tables from PDF documents,
and allows to directly import tables into R in an automated way or by
allowing user to manually select them with a computer mouse thanks to
its integration with \texttt{shiny} (Chang et al. 2023).

\hypertarget{basic-usage}{%
\section{Basic usage}\label{basic-usage}}

We will demonstrate \texttt{tabulapdf} usage by reading tables from a
PDF file created with Quarto (Allaire et al. 2023) and with data
available in R Core Team (2024). The file, included in the package,
contains four tables, with the second and third tables being on the
second page to show a frequent use case that posits some challenges for
table extraction, and it can be accessed from
\href{https://github.com/ropensci/tabulapdf/blob/main/inst/examples/data.pdf}{GitHub}.

The main function, \texttt{extract\_tables()}, mimics the command-line
behavior of Tabula, by extracting tables from a PDF file and, by
default, returns those tables as a list of tibbles in R, where the
column-types are inferred by using \texttt{readr} (Wickham et al. 2019).

The starting point is to load the package and, optionally, to set the
memory allocation for Java:

\begin{Shaded}
\begin{Highlighting}[]
\KeywordTok{library}\NormalTok{(tabulapdf)}

\CommentTok{\# optional: set memory for Java}
\KeywordTok{options}\NormalTok{(}\DataTypeTok{java.parameters =} \StringTok{"{-}Xmx50m"}\NormalTok{)}
\end{Highlighting}
\end{Shaded}

By default, \texttt{extract\_tables()} checks every page for tables
using a detection algorithm and returns all of them:

\begin{Shaded}
\begin{Highlighting}[]
\NormalTok{f \textless{}{-}}\StringTok{ }\KeywordTok{system.file}\NormalTok{(}\StringTok{"examples"}\NormalTok{, }\StringTok{"mtcars.pdf"}\NormalTok{, }\DataTypeTok{package =} \StringTok{"tabulapdf"}\NormalTok{)}
\KeywordTok{extract\_tables}\NormalTok{(f)}
\end{Highlighting}
\end{Shaded}

\begin{verbatim}
## [[1]]
## # A tibble: 5 x 12
##   model          mpg   cyl  disp    hp  drat    wt  qsec    vs    am  gear  carb
##   <chr>        <dbl> <dbl> <dbl> <dbl> <dbl> <dbl> <dbl> <dbl> <dbl> <dbl> <dbl>
## 1 Mazda RX4     21       6   160   110  3.9   2.62  16.5     0     1     4     4
## 2 Mazda RX4 W~  21       6   160   110  3.9   2.88  17.0     0     1     4     4
## 3 Datsun 710    22.8     4   108    93  3.85  2.32  18.6     1     1     4     1
## 4 Hornet 4 Dr~  21.4     6   258   110  3.08  3.21  19.4     1     0     3     1
## 5 Hornet Spor~  18.7     8   360   175  3.15  3.44  17.0     0     0     3     2
## 
## [[2]]
## # A tibble: 1 x 5
##   Sepal.Length                   Sepal.Width    Petal.Length Petal.Width Species
##   <chr>                          <chr>          <chr>        <chr>       <chr>  
## 1 "5.10\r4.90\r4.70\r4.60\r5.00" "3.50\r3.00\r~ "1.40\r1.40~ "0.20\r0.2~ "setos~
## 
## [[3]]
## # A tibble: 5 x 3
##     len supp   dose
##   <dbl> <chr> <dbl>
## 1   4.2 VC      0.5
## 2  11.5 VC      0.5
## 3   7.3 VC      0.5
## 4   5.8 VC      0.5
## 5   6.4 VC      0.5
\end{verbatim}

As you can see for the second table in the output, the result is not
perfect with the default parameters, which is why \texttt{tabulapdf}
provides additional functionality to improve the extraction. In some
cases the extraction should work without additional arguments.

The \texttt{pages} argument allows to select which pages to attempt to
extract tables from:

\begin{Shaded}
\begin{Highlighting}[]
\KeywordTok{extract\_tables}\NormalTok{(f, }\DataTypeTok{pages =} \DecValTok{1}\NormalTok{)}
\end{Highlighting}
\end{Shaded}

\begin{verbatim}
## [[1]]
## # A tibble: 5 x 12
##   model          mpg   cyl  disp    hp  drat    wt  qsec    vs    am  gear  carb
##   <chr>        <dbl> <dbl> <dbl> <dbl> <dbl> <dbl> <dbl> <dbl> <dbl> <dbl> <dbl>
## 1 Mazda RX4     21       6   160   110  3.9   2.62  16.5     0     1     4     4
## 2 Mazda RX4 W~  21       6   160   110  3.9   2.88  17.0     0     1     4     4
## 3 Datsun 710    22.8     4   108    93  3.85  2.32  18.6     1     1     4     1
## 4 Hornet 4 Dr~  21.4     6   258   110  3.08  3.21  19.4     1     0     3     1
## 5 Hornet Spor~  18.7     8   360   175  3.15  3.44  17.0     0     0     3     2
\end{verbatim}

It is possible to specify a remote file, which will be copied to a
temporary directory internally handled by R:

\begin{Shaded}
\begin{Highlighting}[]
\NormalTok{f2 \textless{}{-}}\StringTok{ "https://raw.githubusercontent.com/ropensci/tabulapdf/main/inst/examples/mtcars.pdf"}
\KeywordTok{extract\_tables}\NormalTok{(f2, }\DataTypeTok{pages =} \DecValTok{1}\NormalTok{)}
\end{Highlighting}
\end{Shaded}

\begin{verbatim}
## [[1]]
## # A tibble: 5 x 12
##   model          mpg   cyl  disp    hp  drat    wt  qsec    vs    am  gear  carb
##   <chr>        <dbl> <dbl> <dbl> <dbl> <dbl> <dbl> <dbl> <dbl> <dbl> <dbl> <dbl>
## 1 Mazda RX4     21       6   160   110  3.9   2.62  16.5     0     1     4     4
## 2 Mazda RX4 W~  21       6   160   110  3.9   2.88  17.0     0     1     4     4
## 3 Datsun 710    22.8     4   108    93  3.85  2.32  18.6     1     1     4     1
## 4 Hornet 4 Dr~  21.4     6   258   110  3.08  3.21  19.4     1     0     3     1
## 5 Hornet Spor~  18.7     8   360   175  3.15  3.44  17.0     0     0     3     2
\end{verbatim}

\hypertarget{specifying-the-extraction-method}{%
\section{Specifying the extraction
method}\label{specifying-the-extraction-method}}

For each page, \texttt{extract\_tables()} uses an algorithm to determine
whether it contains one consistent table and then extracts it by using a
spreadsheet-tailored algorithm with the default parameter
\texttt{method\ =\ "lattice"}.

The correct recognition of a table depends on whether the page contains
a table grid. If that is not the case, and the table is a matrix of
cells with values without borders, it might not be able to recognise it.

The same issue appears when multiple tables with different number of
columns are present on the same page. In that case, the parameter
\texttt{method\ =\ "stream"} can be a better option as it will use the
distances between text characters on the page:

\begin{Shaded}
\begin{Highlighting}[]
\CommentTok{\# incorrect}
\KeywordTok{extract\_tables}\NormalTok{(f2, }\DataTypeTok{pages =} \DecValTok{2}\NormalTok{, }\DataTypeTok{method =} \StringTok{"lattice"}\NormalTok{)[[}\DecValTok{1}\NormalTok{]]}
\end{Highlighting}
\end{Shaded}

\begin{verbatim}
## # A tibble: 1 x 5
##   Sepal.Length                   Sepal.Width    Petal.Length Petal.Width Species
##   <chr>                          <chr>          <chr>        <chr>       <chr>  
## 1 "5.10\r4.90\r4.70\r4.60\r5.00" "3.50\r3.00\r~ "1.40\r1.40~ "0.20\r0.2~ "setos~
\end{verbatim}

\begin{Shaded}
\begin{Highlighting}[]
\CommentTok{\# correct}
\KeywordTok{extract\_tables}\NormalTok{(f2, }\DataTypeTok{pages =} \DecValTok{2}\NormalTok{, }\DataTypeTok{method =} \StringTok{"stream"}\NormalTok{)[[}\DecValTok{1}\NormalTok{]]}
\end{Highlighting}
\end{Shaded}

\begin{verbatim}
## # A tibble: 5 x 5
##   Sepal.Length Sepal.Width Petal.Length Petal.Width Species
##          <dbl>       <dbl>        <dbl>       <dbl> <chr>  
## 1          5.1         3.5          1.4         0.2 setosa 
## 2          4.9         3            1.4         0.2 setosa 
## 3          4.7         3.2          1.3         0.2 setosa 
## 4          4.6         3.1          1.5         0.2 setosa 
## 5          5           3.6          1.4         0.2 setosa
\end{verbatim}

\hypertarget{extracting-areas}{%
\section{Extracting areas}\label{extracting-areas}}

\texttt{tabulapdf} uses a table detection algorithm to automatically
identify tables within each page of a PDF. This automatic detection can
be disables with the parameter \texttt{guess\ =\ FALSE} and specifying
an area within each PDF page to extract the table from.

The \texttt{area} argument should be a list either of length equal to
the number of pages specified, allowing the extraction of multiple areas
from one page if the page is specified twice and with two areas
separately:

\begin{Shaded}
\begin{Highlighting}[]
\KeywordTok{extract\_tables}\NormalTok{(}
\NormalTok{  f,}
  \DataTypeTok{pages =} \KeywordTok{c}\NormalTok{(}\DecValTok{2}\NormalTok{, }\DecValTok{2}\NormalTok{),}
  \DataTypeTok{area =} \KeywordTok{list}\NormalTok{(}\KeywordTok{c}\NormalTok{(}\DecValTok{58}\NormalTok{, }\DecValTok{125}\NormalTok{, }\DecValTok{182}\NormalTok{, }\DecValTok{488}\NormalTok{), }\KeywordTok{c}\NormalTok{(}\DecValTok{387}\NormalTok{, }\DecValTok{125}\NormalTok{, }\DecValTok{513}\NormalTok{, }\DecValTok{492}\NormalTok{)),}
  \DataTypeTok{guess =} \OtherTok{FALSE}
\NormalTok{)}
\end{Highlighting}
\end{Shaded}

\begin{verbatim}
## [[1]]
## # A tibble: 5 x 5
##   Sepal.Length Sepal.Width Petal.Length Petal.Width Species
##          <dbl>       <dbl>        <dbl>       <dbl> <chr>  
## 1          5.1         3.5          1.4         0.2 setosa 
## 2          4.9         3            1.4         0.2 setosa 
## 3          4.7         3.2          1.3         0.2 setosa 
## 4          4.6         3.1          1.5         0.2 setosa 
## 5          5           3.6          1.4         0.2 setosa 
## 
## [[2]]
## # A tibble: 5 x 5
##   Sepal.Length Sepal.Width Petal.Length Petal.Width Species  
##          <dbl>       <dbl>        <dbl>       <dbl> <chr>    
## 1          6.7         3            5.2         2.3 virginica
## 2          6.3         2.5          5           1.9 virginica
## 3          6.5         3            5.2         2   virginica
## 4          6.2         3.4          5.4         2.3 virginica
## 5          5.9         3            5.1         1.8 virginica
\end{verbatim}

\hypertarget{interactive-table-extraction}{%
\section{Interactive table
extraction}\label{interactive-table-extraction}}

In addition to the programmatic extraction offered by
\texttt{extract\_tables()}, it is also possible to work interactively
with PDFs. The \texttt{locate\_areas()} function allows to use a
computer mouse to select areas on each page of a PDF, which can then be
used to extract tables:

\begin{figure}[H]
\includegraphics[width=1\linewidth,height=0.3\textheight]{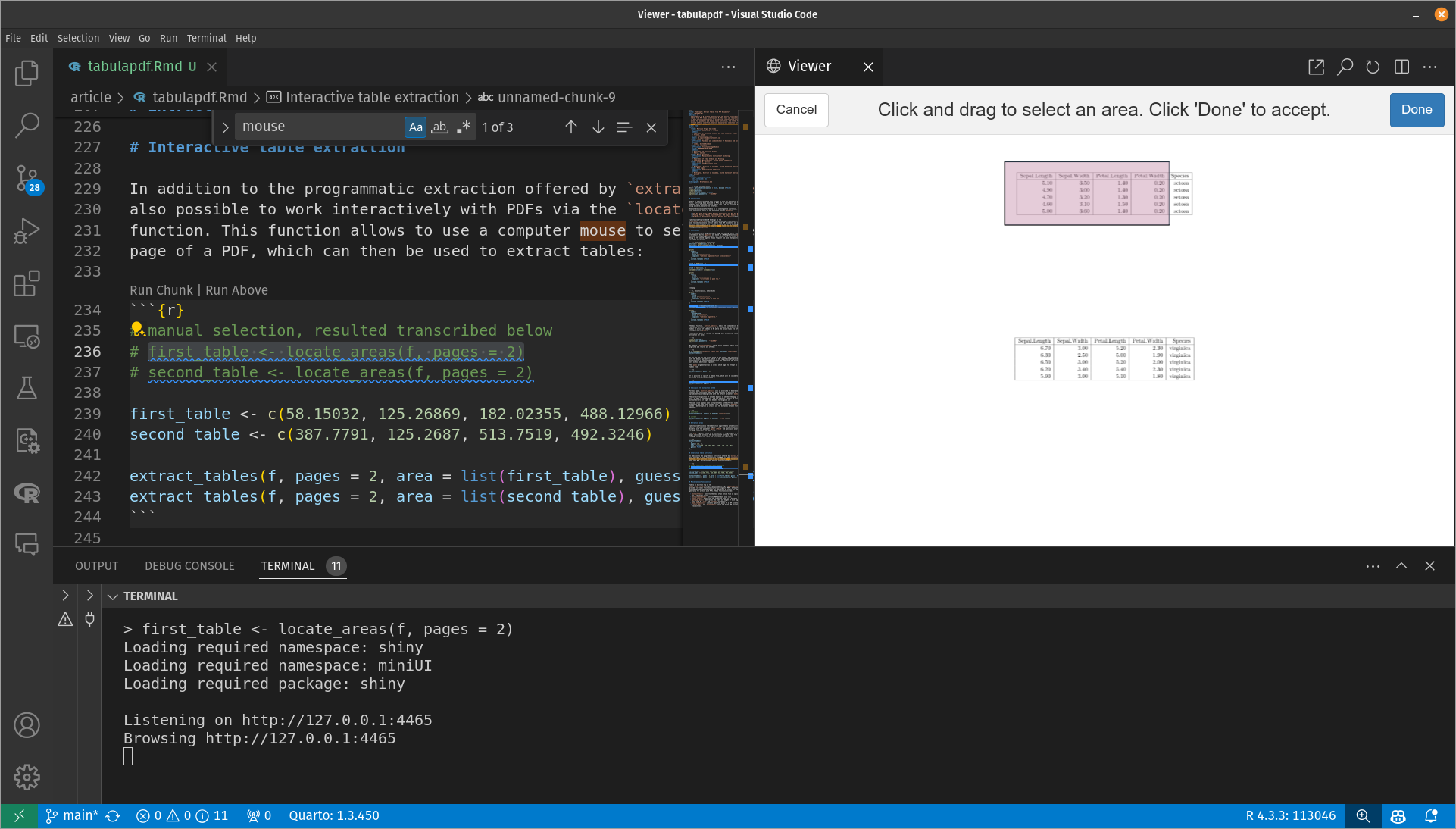} \caption{Selecting areas for table extraction.}\label{fig:unnamed-chunk-7}
\end{figure}

The selection from Figure 1 can be used to extract the tables as
follows:

\begin{Shaded}
\begin{Highlighting}[]
\CommentTok{\# manual selection, result transcribed below}
\CommentTok{\# first\_table \textless{}{-} locate\_areas(f, pages = 2)}
\CommentTok{\# second\_table \textless{}{-} locate\_areas(f, pages = 2)}

\NormalTok{first\_table \textless{}{-}}\StringTok{ }\KeywordTok{c}\NormalTok{(}\FloatTok{58.15032}\NormalTok{, }\FloatTok{125.26869}\NormalTok{, }\FloatTok{182.02355}\NormalTok{, }\FloatTok{488.12966}\NormalTok{)}
\NormalTok{second\_table \textless{}{-}}\StringTok{ }\KeywordTok{c}\NormalTok{(}\FloatTok{387.7791}\NormalTok{, }\FloatTok{125.2687}\NormalTok{, }\FloatTok{513.7519}\NormalTok{, }\FloatTok{492.3246}\NormalTok{)}

\KeywordTok{extract\_tables}\NormalTok{(f, }\DataTypeTok{pages =} \DecValTok{2}\NormalTok{, }\DataTypeTok{area =} \KeywordTok{list}\NormalTok{(first\_table), }\DataTypeTok{guess =} \OtherTok{FALSE}\NormalTok{)}
\end{Highlighting}
\end{Shaded}

\begin{verbatim}
## [[1]]
## # A tibble: 5 x 5
##   Sepal.Length Sepal.Width Petal.Length Petal.Width Species
##          <dbl>       <dbl>        <dbl>       <dbl> <chr>  
## 1          5.1         3.5          1.4         0.2 setosa 
## 2          4.9         3            1.4         0.2 setosa 
## 3          4.7         3.2          1.3         0.2 setosa 
## 4          4.6         3.1          1.5         0.2 setosa 
## 5          5           3.6          1.4         0.2 setosa
\end{verbatim}

\begin{Shaded}
\begin{Highlighting}[]
\KeywordTok{extract\_tables}\NormalTok{(f, }\DataTypeTok{pages =} \DecValTok{2}\NormalTok{, }\DataTypeTok{area =} \KeywordTok{list}\NormalTok{(second\_table), }\DataTypeTok{guess =} \OtherTok{FALSE}\NormalTok{)}
\end{Highlighting}
\end{Shaded}

\begin{verbatim}
## [[1]]
## # A tibble: 5 x 5
##   Sepal.Length Sepal.Width Petal.Length Petal.Width Species  
##          <dbl>       <dbl>        <dbl>       <dbl> <chr>    
## 1          6.7         3            5.2         2.3 virginica
## 2          6.3         2.5          5           1.9 virginica
## 3          6.5         3            5.2         2   virginica
## 4          6.2         3.4          5.4         2.3 virginica
## 5          5.9         3            5.1         1.8 virginica
\end{verbatim}

\hypertarget{use-case-covid-19-treatments-in-italy}{%
\section{Use case: COVID-19 treatments in
Italy}\label{use-case-covid-19-treatments-in-italy}}

It is possible to extract the tables containing the number of
pharmaceutical treatments for hospitalized patients from the
\emph{Monitoraggio Antivirali per COVID-19} (Antiviral Monitoring for
COVID-19) (Agenzia Italiana del Farmaco 2023).

This report features convenient properties:

\begin{enumerate}
\def\labelenumi{\arabic{enumi}.}
\tightlist
\item
  All the links contain ``report\_n'' in the URL, which allows to
  download the files in a straightforward way.
\item
  The table of interest is on page four and keeps its format and
  location in most of the cases.
\end{enumerate}

The following code downloads the HTML page and extracts the links to the
PDF files to be downloaded by using the \texttt{tidyverse} (Wickham et
al. 2019):

\begin{Shaded}
\begin{Highlighting}[]
\KeywordTok{library}\NormalTok{(glue)}
\KeywordTok{library}\NormalTok{(rvest)}
\KeywordTok{library}\NormalTok{(stringr)}
\KeywordTok{library}\NormalTok{(dplyr)}

\NormalTok{url \textless{}{-}}\StringTok{ "https://www.aifa.gov.it/uso{-}degli{-}antivirali{-}orali{-}per{-}covid{-}19"}

\NormalTok{dout \textless{}{-}}\StringTok{ "covid\_reports"}
\KeywordTok{try}\NormalTok{(}\KeywordTok{dir.create}\NormalTok{(}\StringTok{"covid\_reports"}\NormalTok{), }\DataTypeTok{silent =} \OtherTok{TRUE}\NormalTok{)}

\NormalTok{finp \textless{}{-}}\StringTok{ }\KeywordTok{glue}\NormalTok{(}\StringTok{"\{ dout \}/links.txt"}\NormalTok{)}

\ControlFlowTok{if}\NormalTok{ (}\OperatorTok{!}\KeywordTok{file.exists}\NormalTok{(finp)) \{}
\NormalTok{  links \textless{}{-}}\StringTok{ }\KeywordTok{read\_html}\NormalTok{(url) }\OperatorTok{\%\textgreater{}\%}
\StringTok{    }\KeywordTok{html\_nodes}\NormalTok{(}\StringTok{"a"}\NormalTok{) }\OperatorTok{\%\textgreater{}\%}
\StringTok{    }\KeywordTok{html\_attr}\NormalTok{(}\StringTok{"href"}\NormalTok{) }\OperatorTok{\%\textgreater{}\%}
\StringTok{    }\KeywordTok{str\_subset}\NormalTok{(}\StringTok{"report\_n"}\NormalTok{)}
  \KeywordTok{writeLines}\NormalTok{(links, finp)}
\NormalTok{\} }\ControlFlowTok{else}\NormalTok{ \{}
\NormalTok{  links \textless{}{-}}\StringTok{ }\KeywordTok{readLines}\NormalTok{(finp)}
\NormalTok{\}}

\NormalTok{links \textless{}{-}}\StringTok{ }\KeywordTok{glue}\NormalTok{(}\StringTok{"https://www.aifa.gov.it\{ links \}"}\NormalTok{)}
\end{Highlighting}
\end{Shaded}

It is not the case for this data, but combined cells in the header can
difficult the extraction process. It is possible to select the figures
only and set the \texttt{col\_names} argument to \texttt{FALSE} to avoid
this issue.

The following code downloads and tidies the table for the first report:

\begin{Shaded}
\begin{Highlighting}[]
\NormalTok{finp \textless{}{-}}\StringTok{ }\KeywordTok{glue}\NormalTok{(}\StringTok{"covid\_reports/\{ basename(links[1]) \}"}\NormalTok{)}

\ControlFlowTok{if}\NormalTok{ (}\OperatorTok{!}\KeywordTok{file.exists}\NormalTok{(finp)) \{}
  \KeywordTok{try}\NormalTok{(}\KeywordTok{download.file}\NormalTok{(links[}\DecValTok{1}\NormalTok{], }\DataTypeTok{destfile =}\NormalTok{ finp, }\DataTypeTok{quiet =} \OtherTok{TRUE}\NormalTok{))}
\NormalTok{\}}

\CommentTok{\# locate\_areas(finp, pages = 4)}

\NormalTok{report1 \textless{}{-}}\StringTok{ }\KeywordTok{extract\_tables}\NormalTok{(finp,}
  \DataTypeTok{pages =} \DecValTok{4}\NormalTok{, }\DataTypeTok{guess =} \OtherTok{FALSE}\NormalTok{, }\DataTypeTok{col\_names =} \OtherTok{FALSE}\NormalTok{,}
  \DataTypeTok{area =} \KeywordTok{list}\NormalTok{(}\KeywordTok{c}\NormalTok{(}\FloatTok{140.75}\NormalTok{, }\FloatTok{88.14}\NormalTok{, }\FloatTok{374.17}\NormalTok{, }\FloatTok{318.93}\NormalTok{))}
\NormalTok{)}

\NormalTok{report1 \textless{}{-}}\StringTok{ }\NormalTok{report1[[}\DecValTok{1}\NormalTok{]] }\OperatorTok{\%\textgreater{}\%}
\StringTok{  }\KeywordTok{rename}\NormalTok{(}\DataTypeTok{region =}\NormalTok{ X1, }\DataTypeTok{treatments =}\NormalTok{ X2, }\DataTypeTok{pct\_increase =}\NormalTok{ X3) }\OperatorTok{\%\textgreater{}\%}
\StringTok{  }\KeywordTok{mutate}\NormalTok{(}
    \DataTypeTok{treatments =} \KeywordTok{as.numeric}\NormalTok{(}\KeywordTok{str\_replace}\NormalTok{(treatments, }\StringTok{"}\CharTok{\textbackslash{}\textbackslash{}}\StringTok{."}\NormalTok{, }\StringTok{""}\NormalTok{)),}
    \DataTypeTok{pct\_increase =}\NormalTok{ pct\_increase }\OperatorTok{\%\textgreater{}\%}
\StringTok{      }\KeywordTok{str\_replace}\NormalTok{(}\StringTok{","}\NormalTok{, }\StringTok{"."}\NormalTok{) }\OperatorTok{\%\textgreater{}\%}
\StringTok{      }\KeywordTok{str\_replace}\NormalTok{(}\StringTok{"\%"}\NormalTok{, }\StringTok{""}\NormalTok{) }\OperatorTok{\%\textgreater{}\%}
\StringTok{      }\KeywordTok{as.numeric}\NormalTok{(.) }\OperatorTok{/}\StringTok{ }\DecValTok{100}\NormalTok{,}
    \DataTypeTok{date =}\NormalTok{ finp }\OperatorTok{\%\textgreater{}\%}
\StringTok{      }\KeywordTok{basename}\NormalTok{() }\OperatorTok{\%\textgreater{}\%}
\StringTok{      }\KeywordTok{str\_replace}\NormalTok{(}\StringTok{".*antivirali\_"}\NormalTok{, }\StringTok{""}\NormalTok{) }\OperatorTok{\%\textgreater{}\%}
\StringTok{      }\KeywordTok{str\_replace}\NormalTok{(}\StringTok{"}\CharTok{\textbackslash{}\textbackslash{}}\StringTok{.pdf"}\NormalTok{, }\StringTok{""}\NormalTok{) }\OperatorTok{\%\textgreater{}\%}
\StringTok{      }\KeywordTok{as.Date}\NormalTok{(}\DataTypeTok{format =} \StringTok{"\%d.\%m.\%Y"}\NormalTok{)}
\NormalTok{  )}

\NormalTok{report1}
\end{Highlighting}
\end{Shaded}

\begin{verbatim}
## # A tibble: 22 x 4
##    region                treatments pct_increase date      
##    <chr>                      <dbl>        <dbl> <date>    
##  1 Abruzzo                     2343        0.03  2022-01-21
##  2 Basilicata                   927        0.012 2022-01-21
##  3 Calabria                    1797        0.023 2022-01-21
##  4 Campania                    3289        0.041 2022-01-21
##  5 Emilia Romagna              7945        0.1   2022-01-21
##  6 Friuli Venezia Giulia       1063        0.013 2022-01-21
##  7 Lazio                      11206        0.141 2022-01-21
##  8 Liguria                     5332        0.067 2022-01-21
##  9 Lombardia                  12089        0.153 2022-01-21
## 10 Marche                      3739        0.047 2022-01-21
## # i 12 more rows
\end{verbatim}

\hypertarget{miscellaneous-functionality}{%
\section{Miscellaneous
functionality}\label{miscellaneous-functionality}}

Tabula is built on top of the \href{https://pdfbox.apache.org/}{Java
PDFBox library} (Litchfield 2024), which provides low-level
functionality for working with PDFs. A few of these tools are exposed
through \texttt{tabulapdf}, as they might be useful for debugging or
generally for working with PDFs. These functions include:

\begin{itemize}
\tightlist
\item
  \texttt{extract\_text()} converts the text of an entire file or
  specified pages into an R character vector.
\item
  \texttt{extract\_metadata()} extracts PDF metadata as a list.
\item
  \texttt{get\_n\_pages()} determines the number of pages in a document.
\item
  \texttt{get\_page\_dims()} determines the width and height of each
  page in pt (the unit used by \texttt{area} and \texttt{columns}
  arguments).
\item
  \texttt{make\_thumbnails()} converts specified pages of a PDF file to
  image files.
\item
  \texttt{split\_pdf()} and \texttt{merge\_pdfs()} split and merge PDF
  documents, respectively.
\end{itemize}

\hypertarget{conclusion}{%
\section{Conclusion}\label{conclusion}}

\texttt{tabulapdf} brings a powerful and flexible tool to R users by
integrating the Tabula Java library into R for PDF table extraction. It
extends beyond automation by offering users manual extraction options
through a Shiny interface. By simplifying access to structured data
hidden within PDFs, this package aims to help data extraction in fields
such as investigative journalism, academic research, and other areas
requiring accurate and efficient data retrieval from fixed-layout
documents.

\hypertarget{acknowledgements}{%
\section{Acknowledgements}\label{acknowledgements}}

The underlying Tabula Java library has received support from multiple
anonymous donors on Open Collective, the Knight Foundation, the Mozilla
Foundation, ProPublic, La Nación, The New York Times and the
Shuttleworth Foundation.

The rOpenSci software review process contributed to largely improve
tabulapdf with the feedback from David Gohel and Lincoln Mullen.

\hypertarget{references}{%
\section*{References}\label{references}}
\addcontentsline{toc}{section}{References}

\hypertarget{refs}{}
\begin{cslreferences}
  
\leavevmode\hypertarget{ref-agenzia}{}%
Agenzia Italiana del Farmaco. 2023. ``Uso Degli Antivirali Per
COVID-19.''
\url{https://www.aifa.gov.it/uso-degli-antivirali-orali-per-covid-19}.

\leavevmode\hypertarget{ref-quarto}{}%
Allaire, JJ, Christopher Dervieux, Carlos Schneider, Charles Teague, and
Yihui Xie. 2023. \emph{`Quarto`: Open Source Tools for Scientific and
Technical Publishing}. \url{https://CRAN.R-project.org/package=shiny}.

\leavevmode\hypertarget{ref-tabula}{}%
Aristaran, Manuel. 2018. \emph{Tabula: A Tool for Liberating Data Tables
Trapped Inside Pdf Files}. \url{https://tabula.technology/}.

\leavevmode\hypertarget{ref-shiny}{}%
Chang, Winston, Joe Cheng, JJ Allaire, Carson Sievert, Barret Schloerke,
Yihui Xie, Jeff Allen, Jonathan McPherson, Alan Dipert, and Barbara
Borges. 2023. \emph{`Shiny`: Web Application Framework for R}.
\url{https://CRAN.R-project.org/package=shiny}.

\leavevmode\hypertarget{ref-dataversusinformation}{}%
James C. King. 2010. ``Data Versus Information.'' \emph{Inside PDF}.
\url{http://web.archive.org/web/20111221044202/http://blogs.adobe.com/insidepdf/2010/11/my-pdf-hammer.html}.

\leavevmode\hypertarget{ref-pdfhammer}{}%
---------. 2011. ``My PDF Hammer (Revision).'' \emph{Inside PDF}.
\url{http://web.archive.org/web/20111223072718/http://blogs.adobe.com/insidepdf/}.

\leavevmode\hypertarget{ref-apachepdfbox}{}%
Litchfield, Ben. 2024. \emph{Apache Pdfbox: A Java Pdf Library}.
\url{https://pdfbox.apache.org/index.html/}.

\leavevmode\hypertarget{ref-updatingdollars}{}%
Merrill, Jeremy B. 2013. ``Heart of Nerd Darkness: Why Updating Dollars
for Docs Was so Difficult.'' \emph{ProPublica}.
\url{https://www.propublica.org/nerds/heart-of-nerd-darkness-why-dollars-for-docs-was-so-difficult}.

\leavevmode\hypertarget{ref-base}{}%
R Core Team. 2024. \emph{`R`: A Language and Environment for Statistical
Computing}. Vienna, Austria: R Foundation for Statistical Computing.
\url{https://www.R-project.org/}.

\leavevmode\hypertarget{ref-tidyverse}{}%
Wickham, Hadley, Mara Averick, Jennifer Bryan, Winston Chang, Lucy
D'Agostino McGowan, Romain François, Garrett Grolemund, et al. 2019.
``Welcome to the `Tidyverse`.'' \emph{Journal of Open Source Software} 4
(43): 1686. \url{https://doi.org/10.21105/joss.01686}.
\end{cslreferences}


\end{document}